# *Hole doping in compositionally complex correlated oxide enables tunable exchange biasing*

Alessandro R. Mazza,[1,2,*] Elizabeth Skoropata,[1] Jason Lapano,[1] Michael A. Chilcote[1], Cameron Jorgensen,[3] Nan Tang,[3] Zheng Gai,[4] John Singleton,[5] Matthew J. Brahlek,[1] Dustin A. Gilbert,[3] Thomas Z. Ward[1,*]

[1]Materials Science and Technology Division, Oak Ridge National Laboratory, Oak Ridge, Tennessee 37831, USA

[2]Center for Integrated Nanotechnologies, Los Alamos National Laboratory, Los Alamos, NM

[3]Department of Materials Science and Engineering, University of Tennessee, Knoxville, Tennessee 37996-4545, USA

[4]Center for Nanophase Materials Sciences, Oak Ridge National Laboratory, Oak Ridge, Tennessee 37831, USA

[5]National High Magnetic Field Laboratory, Los Alamos National Laboratory, Los Alamos, NM 87545, USA

[*]email: armazza@lanl.gov, wardtz@ornl.gov

Abstract

Magnetic interfaces and the phenomena arising from them drive both the design of modern spintronics and fundamental research. Recently, it was revealed that through designing magnetic frustration in configurationally complex entropy stabilized oxides, exchange bias can occur in structurally single crystal films. This eliminates the need for complex heterostructures and nanocomposites in the design and control of magnetic response phenomena. In this work, we demonstrate through hole doping of a high entropy perovskite oxide that tuning of magnetic responses can be achieved. With detailed magnetometry, we show magnetic coupling exhibiting a variety of magnetic responses including exchange bias and antiferromagnetic spin reversal in the entropy stabilized $ABO_3$ perovskite oxide $La_{1-x}Sr_x(Cr_{0.2}Mn_{0.2}Fe_{0.2}Co_{0.2}Ni_{0.2})O_3$ family. We find that manipulation of the *A*-site charge state can be used to balance magnetic phase compositions and coupling responses. This allows for the creation of highly tunable exchange bias responses. In the low Sr doping regime, a spin frustrated region arising at the antiferromagnetic phase boundary is shown to directly couple to the antiferromagnetic moments of the film and emerges as the dominant mechanism, leading to a vertical shift of magnetization loops in response to field biasing. At higher concentrations, direct coupling of antiferromagnetic and ferromagnetic regions is observed. This tunability of magnetic coupling is discussed within the context of these three competing magnetic phases, revealing critical features in designing exchange bias through exploiting spin frustration and disorder in high entropy oxides.

Introduction

Exchange bias and other magnetic responses resulting from coupled interaction of ferromagnetic (FM) and antiferromagnetic (AFM) constituents, have garnered a continuing interest in both fundamental and applied research. Fundamentally, for example, interfacial coupling [1,2], training effects [3,4], reversal mechanisms [5], and different loop asymmetries [6,7] resulting from field biasing are all of interest. Functionally, as the number of applications and demand for complex magnetic systems grows, the use of artificial heterostructures and nanocomposites is flourishing in fields from quantum computing [8–10] to pharmacology [11]. However, the limitation of nanocomposites is often the high degree of disorder due to roughness, inhomogeneous stoichiometry, size effects and symmetry, which each can diminish the magnetic moment and affect the coupling of magnetic phases. Similarly, artificial heterostructures have limitations with small growth windows and ranges in compositions for which they can be precisely synthesized. By combining the single crystal uniformity of heterostructures with the range of compositions available in the synthesis of nanocomposites, an unprecedented control over magnetic phase and magnetic response should be possible. As a mechanism to such control, we explore the newly emerging entropy stabilized oxide materials class, which host a wide variety of magnetic microstates in uniform single crystal films [12,13].

Entropy-stabilized oxides [14], which quench in a random compositional distribution of atoms on a uniform lattice, have been shown to stabilize in a number of structures, including spinel [15–18], rocksalt [13,19], Ruddlesden-Popper [20,21], and perovskite [22–25]. Within the perovskite class of materials, interesting physics revolving around the magnetic spin and exchange disorder inherent to a configurationally mixed B-site has resulted in phase competition enabling a surprising monolithic exchange bias (EB) [26]. This La($Cr_{0.2}Mn_{0.2}Fe_{0.2}Co_{0.2}Ni_{0.2}$)$O_3$ was a demonstration of two fundamental ideas: (1) the average exchange value dictates the magnetic order parameter and (2) magnetic exchange and spin disorder lead to magnetic phase competition in these structurally perfect systems. The latter is an enticing result not only in magnetic frustration but also in EB applications where precise control of the magnetic response is a necessity.

Furthermore, the demonstrated FM/AFM competition appears to exhibit a clear third magnetic phase, likely glassy in nature, which arises from the magnetic frustration and disorder in the system [27]. These results further garner interest in the EB response as recent studies [28] suggest a close relationship between glassy dynamics and exchange bias through AFM/FM/spin-glass interactions. Expanding on this idea, $La_{1-x}Sr_x(Cr_{0.2}Mn_{0.2}Fe_{0.2}Co_{0.2}Ni_{0.2})O_3$ (LS5BO) is particularly promising in exploring tunable magnetic phenomena, as there is precise control over phase competition between AFM and FM components through hole doping. If considering the local effects of hole doping, it is instructive to investigate the ternary parents, which each exhibit either a suppressed $T_N$ [29–31] or a transition from AFM to FM ordering [32–34] with increasing Sr. It has been shown that in this series the 0% Sr sample exhibits a large vertical shift of the loop upon field biasing [26]. However, the remaining Sr > 0% samples have not been explored. Given the role of charge doping in manipulating the magnetic phases of the parent oxides, LS5BO is an astounding candidate to investigate the role of hole doping in manipulating magnetic phase competition and interaction.

In this work, we investigate the magnetic response to field biasing of LS5BO as a function of hole doping. With detailed magnetometry we show magnetic phase disorder driven pinning (vertical shift), traditional EB (horizontal shift), and antiferromagnetic spin reversal (coercivity enhancement) coupling phenomena entirely dependent on Sr fraction x. The observed magnetic anisotropy is understood within the context of competing magnetic phases and connected to the lattice anisotropy of the strained films. An important feature of the EB response is the spin frustration of the system which gives rise to a glassy region of uncompensated moments which is the dominating feature of low-doped samples' magnetic responses. However, the dominant biased response shifts with x, a result which is connected to increasing FM character of the film resulting in direct AFM and FM coupling. This provides the first example of fine-tuned control of magnetic response – and interaction of AFM and FM regions in samples - to field biasing in entropy stabilized oxides.

Results and Discussion

LS5BO can be characterized by magnetic frustration linked to competing AFM and FM phases inherent to its configurational disorder. This is largely driven by the magnetic exchange and spin disorder on the positionally ideal single crystal samples [26]. The parent L5BO (x = 0) has an affinity for AFM while maintaining small FM pockets. As the Sr concentration increases, these FM pockets become more prevalent in the film. This is enabled by opening of double exchange pathways as the valence of, for example, Mn and Co increase towards 4+. As a result, Sr doping increases the total volume fraction and the magnitude of the ferromagnetic moment [27]. This increase in robust FM comes with the loss of a glassy, soft magnetic feature observed primarily in Sr dopings of under 10%. While magnetic phase was shown to vary with x, the interaction between the order types, and how they compete cannot be inferred from traditional magnetometry alone. For this we turn to field biased measurements. Field bias measurements can help to elucidate the relative energy scales of the effective Zeeman energy of a FM ($E_{FM}$), anisotropy energy of an AFM ($E_{AFM}$), and AFM/FM interfacial energy ($E_{Int}$). The resulting EB, pinning, or AFM spin reversal implies the energy scale and coupling of AFM and FM components in LS5BO films. $E_{Int}$ in this generalization may be one of the most interesting components of this interaction, as a glassy magnetic region at the AFM/FM interface is believed to emerge in LS5BO [27]. The ability to tune such a region gives rise to immediate applied interest as the uncompensated spin glass can couple to the AFM which yields a magnetic response which can far exceed standard heterostructured systems [28]. Figure 1 generally summarizes how the energy scales relevant to EB exhibit different magnetic responses in magnetic systems. The bottom panel of Fig. 1 shows the expected change in loop shape as the relative energies of $E_{FM}$, $E_{AFM}$, and $E_{Int}$ change. As $E_{FM}$ increases, first approaching and then surpassing $E_{AFM}$, the magnetization response changes from displaying a vertical offset to displaying the horizontal shift characteristic of traditional EB. EB results from true coupling of FM and AFM regions, unlike a vertical shift that can occur from the pinning of uncompensated spins at phase boundaries by an AFM. Generally, EB can be thought of as being driven by the manifestation of an energetic cost to the switching of a ferromagnet. This contrasts with pinning, which is seen when a soft magnetic component has proximity to a robust AFM - thereby "pinning" interfacial spins to the field bias direction resulting in a surplus magnetization. With sufficiently large relative $E_{FM}$ and $E_{Int}$, the spins of the AFM pockets can be reversed as well, manifesting as an enhancement to the coercivity in comparison to measurements done without bias. This explanation is limited to the very local interaction of magnetic phases in materials and the mechanism for exchange bias in disordered materials is still a subject of debate. In our samples, for example, the local interfaces can have any number of orientations between phases and the mechanisms described here, without some preferred orientation, could lead to an averaging effect which is unobservable. The assumption made here is that the phases each have preferred crystallographic directions (relative to the crystal orientation of the film) which avoid this "averaging out" of the biased responses of the global system. This contrasts with nanocomposites where crystallites themselves are randomly orientated and any anisotropy along a crystal axis cannot be measured. Each of these magnetic responses to field bias are represented also in the right panels of Fig. 1, which show a cartoon representation of spins at an AFM/FM interface after field reversal. In this representation the interface is shown to be sharp though, as discussed later in the text, the interface may be dominant in the biased response as was seen in recent reports of coupling between AFM/spin glass magnetic layers [28]. This is in some contrast to heterostructured systems such as $LaMnO_3/LaFeO_3$ and $LaMnO_3$ where a sharp interface exists between two otherwise magnetically isolated phases [35,36].

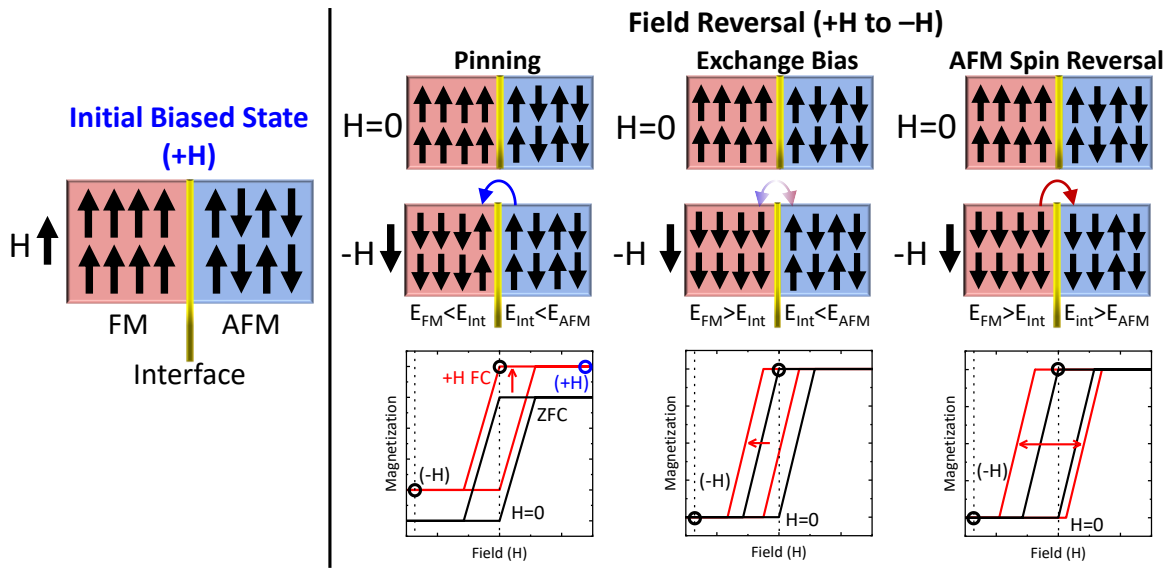

Fig. 1 The role of coexisting Zeeman energy of ferromagnet ($E_{FM}$), anisotropy energy of antiferromagnetic ($E_{AFM}$), and interface energy ($E_{int}$) in determining functional magnetic response under field cooling and field reversal. The initial biased state is shown on the left (+H) and the response of spins in the FM (red) and AFM (blue) to a switch in field direction (through H=0 to -H as labeled) are connected to their manifestation in the changes in loop shapes. The biased state is labeled as (+H) for a saturating +H field and the field reversal is labeled as (-H) for a saturating –H field. The relative energies of the magnetic phase of coexisting domains and domain walls dictate field biasing's effect on magnetic loop offset and hysteresis. For comparison zero field cooled (ZFC) loops are compared directly to field cooled (FC) loops.

Considering the interplay of AFM/FM bonds in the LS5BO system we perform field biased measurements ± 7 T for both the in-plane (IP) and out-of-plane (OOP) conditions. Samples are each ~ 50 nm thick and an extensive summary of their structural properties and growth can be found in Refs. [12,23,27] and discussed in the methods section. Field cooled measurements collected at 2.1 K with the field applied in-plane are summarized for all x in Fig. 2(a) where the black colored loops are +7 T field biased, and the red are -7 T field biased. Clearly in each case a vertical shift is observed but upon subtraction of this vertical shift, EB also appears. These results are importantly mediated by $E_{Int}$, which is a measure of the resistance of the interface in allowing coupling of AFM and FM in the film. In the OOP direction (Fig. 2(b)), EB is more apparent for x=0.3, 0.5 and diminished in lower dopings. The OOP direction was found to be the axis which exhibits a strong anisotropic (mixed soft/hard) loop shapes in the x=0, 0.1 cases and is the easy axis of the FM for all x in tensile strained films, suggesting $E_{FM}$ would be strongest in this direction [27]. $E_{FM}$ and $E_{AFM}$ both are subject to change due to field orientation, evident from the anisotropy observed between IP/OOP directions, and as a function of hole doping. $E_{Int}$ appears tunable with Sr, with a third glassy magnetic region associated with the AFM/FM interface emerging most clearly in 0% and 10% samples and largely dissipating at higher concentrations. This component is likely responsible for the vertical pinning seen in the EB results and is closely tied to $E_{Int}$ for these dopings. As a source of uncompensated spins in vertical pinning, the mechanism which drives the biased response sheds light into the broader magnetic behavior of this component. A newly suggested mechanism for exchange bias [28] is consistent with this region being spin glass, arising from a degenerate landscape that is known to exist in L5BO. It is this spin glass/AFM coupling which may effectively be tuned by Sr doping as the region appears to collapse with increasing Sr, where the anisotropic loop shape attributed to the uncompensated moments in this region disappear for x>0.1 [27].

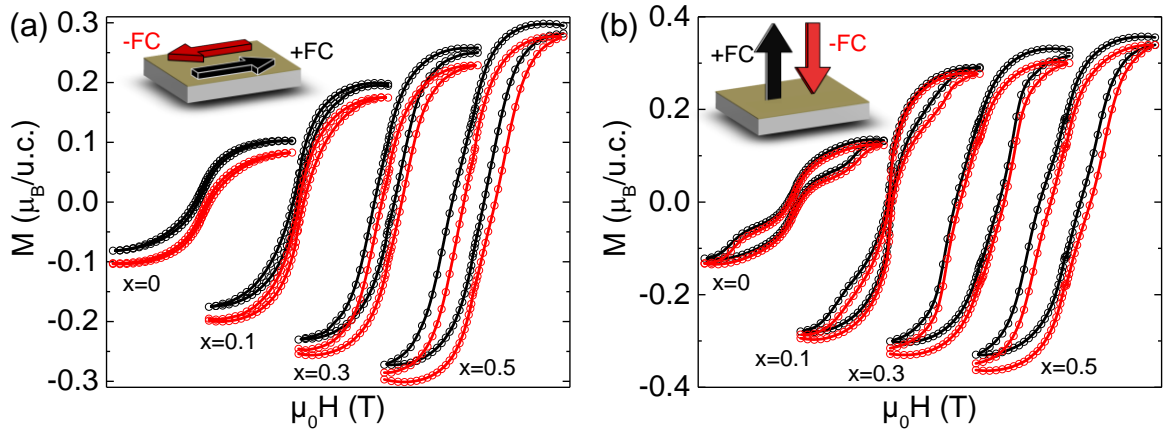

Fig. 2 Magnetic responses of films after field cooling to 2.1 K from 300 K under ±7 T fields applied (a) in-plane and (b) out-of-plane directions.

Results of the field bias measurements (± 7 T bias) at 2.1 K for both IP and OOP directions are summarized in Fig. 3. Not included in this summary is the enhancement of the coercivity observed with increased Sr doping. This enhancement was also observed in unbiased loops [27] and supports the interpretation that there is an increase in FM cluster density and domain boundaries between FM/AFM portions of the film which allow for direct coupling between the two magnetic phases. In the OOP direction, the EB is absent in both x = 0, 0.1 and diminished in the x=0.3, 0.5 cases. If we investigate more closely, this reveals the delicate balance of the relevant energies in the system. In this discussion, it is important to note the assumption that the exchange energy of the interface ($E_{Int}$) is isotropic. This is generally true in, for example, nanocomposites despite magnetic and crystalline anisotropy [37,38]. In the x=0, 0.1 cases there are two scenarios which allow the disappearance of EB OOP: if we assume EB implies $E_{FM}>E_{Int}$ and $E_{AFM}>E_{Int}$, the absence of EB in the OOP direction can be due to either (1) a reduction of $E_{FM}$ such that $E_{FM}<E_{Int}$, resulting in a pinning response or (2) a decrease in both $E_{FM}$ and $E_{AFM}$ in the OOP direction. Since there is a small reduction to the pinning response OOP, the second scenario with a reduction in both $E_{AFM}$ and $E_{FM}$ seems the likely consequence of the films' magnetic anisotropy. This leads to $E_{FM} < E_{Int}$ and, if $E_{AFM} < E_{FM}$, spins are free to rotate yet unable to overcome the interfacial energy barrier to have a response to field biasing. With diminished $E_{AFM}$, the uncompensated spins of the glassy region can freely rotate which allows the observed contribution to the uncoupled soft/hard loops (Fig. 2(b)) shapes seen at small Sr concentrations and especially prevalent in the 10% Sr films, where the soft response is believed to be dominated by these free spins. This is likely tied to magneto-crystalline effects which drive anisotropy in both FM and AFM materials [39].

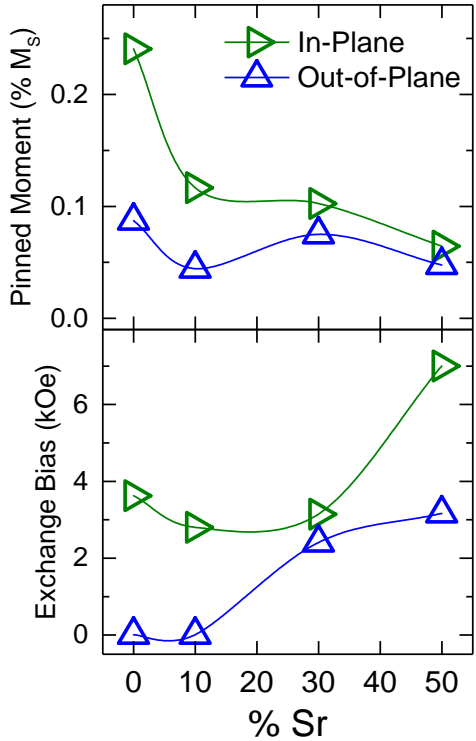

Fig. 3 Summary of the field biased measurements at 2.1K for all Sr concentrations for in-plane and out-of-plane geometries.

The higher hole doping samples exhibit a different mechanism for anisotropy in the magnetic response. In the x = 0.3 case EB is also diminished OOP as compared to IP but to a very small degree as compared to x = 0, 0.1. However, the dramatic increase in EB in the OOP as compared to 0% and 10% Sr samples supports the idea that the FM clusters become more robust and the regions containing uncompensated spins at the edges of AFM regions begin to dissipate with hole doping and allow direct coupling of AFM/FM regions of the film. For the x = 0.5 film, the change in EB is not the complete story. IP the x = 0.5 sample shows a large coercivity enhancement in the FC loops compared to ZFC loops. At 2.1 K for the x = 0.5 case the coercivity enhancement is a factor of nearly 2 higher in comparison to ZFC measurements by AFM spin reversal (AFR) seen in Fig. 4(a). As we see AFR in the IP direction, we must have $E_{AFM} < E_{Int}$, $E_{FM} > E_{Int}$ and (therefore) $E_{AFM} < E_{FM}$. However, in the OOP direction (Fig. 4(b)), along with the diminished EB, the AFM spin reversal disappears. This suggests again the case of $E_{AFM} < E_{Int}$ and $E_{FM} < E_{Int}$ coupled with $E_{FM} > E_{AFM}$, implying only a drop in $E_{FM}$ in the OOP direction as we cannot directly observe if $E_{AFM}$ varies. However, given the direct relationship of $E_{AFM}$ and $E_{FM}$ in each of the other x it is believed that they are both largest IP with a reduction in energy for both appearing in the OOP direction. This is consistent with their lattice anisotropy, which is along the OOP direction, being the driving force in the magnetic anisotropy of the films via magneto-crystalline effects. These changes are concatenated with a large increase in coercivity, mentioned above. We comment here that this change in coercivity is so dramatic that, in Fig. 2, the biased loops do not fully close, implying they are on a minor loop. This large enhancement of the coercivity merits future study towards the potential for the frustrated magnetic nature of HEOs yielding extremely hard magnetic behaviors.

A qualitative summary of the evolution of the magnetic response of films and how the balance of energies giving rise to EB changes as a function of x is shown in Fig. 5. As an important note, this is a qualitative representation of the changes observed as a function of hole doping and therefore the magnitudes of the changes of each of the energies in Fig. 5 should not be compared. It is the interplay of these energies that we are able to probe. Remarkably, through this range of hole doping, we have unique and fine control of the magnetic response of the films. However, given this fine-tuned control, regions of the film still seem to lag. For example, as shown in Fig. 3, the x = 0.3 and 0.5 cases still show a vertical shift despite being dominated by EB and AFM spin reversal. This suggests that while the films are structurally uniform, the magnetic structure is remarkably disordered with a clear mix of FM, AFM and a glassy region of uncompensated spins. This spin frustration at the interface arising from magnetic exchange and spin disorder is unique in its contribution to understanding the mechanisms for EB and possible avenues toward designing and manipulating magnetic dynamics in high entropy oxides. As Sr concentration increases, the AFM/FM interface containing a region of uncompensated spins collapses and the vertical pinning appears to weaken. This is likely due to the increasing FM nature with Sr, which can be connected to local double exchange pathways being created by a shift in the average charge of the perovskite B-site [27]. It is this local and significant increase in FM character that allows direct coupling between AFM and FM phases in the films. The control over EB exhibited here establishes the materials design strategy exploiting magnetic frustration in high entropy oxides as a viable and intriguing path towards broad technological application, where disorder of magnetic exchange interactions and spins can be tuned towards the desired functionality. The monolithic EB response in a single crystal rivals the applicability of nanocomposites in devices. In the light of the proposed glassy region, which plays a key role in the coupling mechanism, this may be extrapolated into a larger picture to explore dynamic responses within the materials class. This is especially pertinent as we demonstrate the ability to tune the AFM phase boundary, which hosts this glassy region of uncompensated spins. This result guides future exploration of phase competition in entropy stabilized oxides and informs the importance of a robust FM phase, enhanced in LS5BO by double exchange, in designing single-crystal materials exhibiting EB.

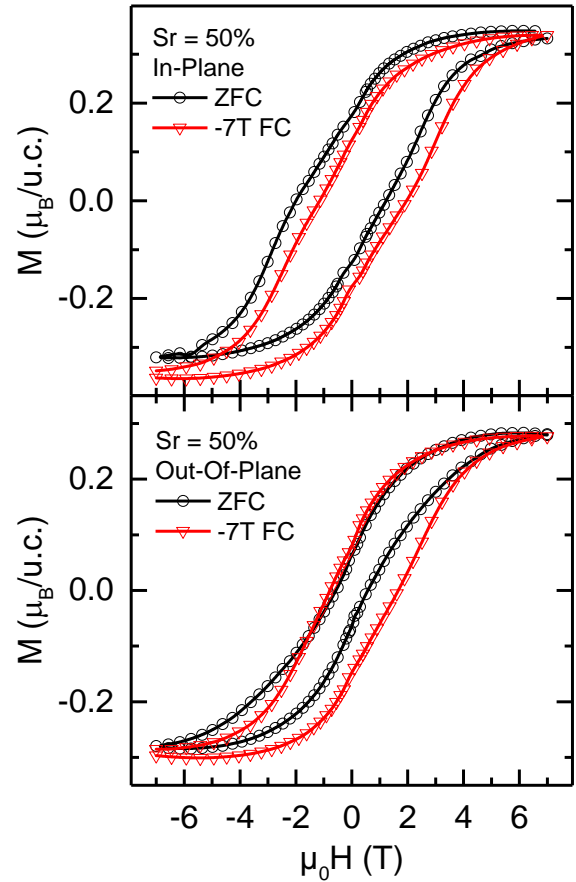

Fig. 4 AFM spin reversal response is present with field applied in-plane (top) but is absent in the out-of-plane geometry(bottom) for the x = 0.5 at 2.1 K.

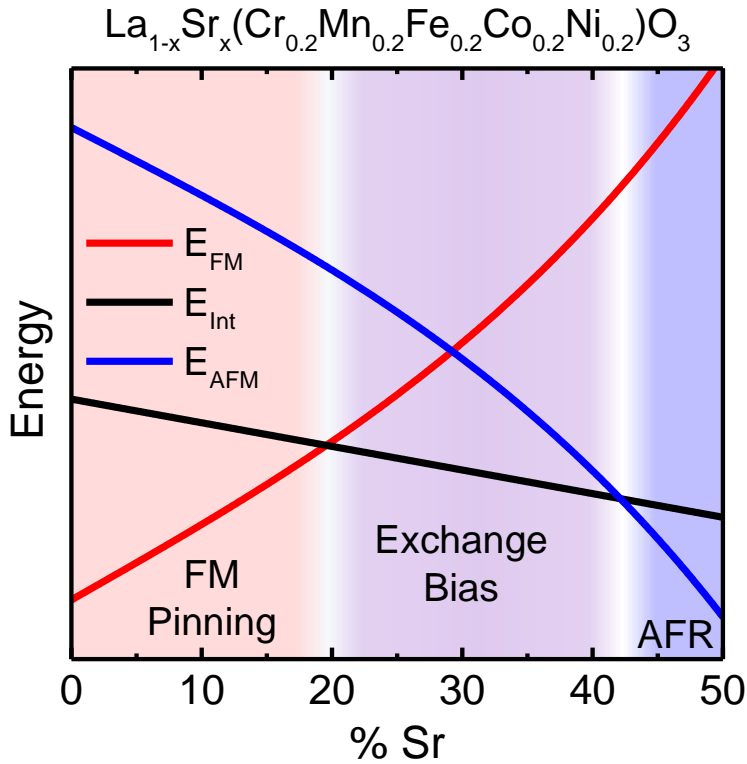

Fig. 5 Phase diagram of dominant magnetic response changes with Sr concentration, x. The energies are qualitatively drawn to show evolution of dominant functional response associated with pinning, traditional exchange bias, and antiferromagnetic spin reversal (AFR). The magnitudes of these changes are not comparable.

Methods

LS5BO films were synthesized using pulsed laser deposition as described in [27]. To briefly summarize, samples are synthesized using pulsed laser deposition from stoichiometric single phase ceramic targets. Each of the films are grown on SrTiO$_3$ substrates using a KrF excimer laser with a laser fluence of 0.85 J/cm$^2$ and a pulse rate of 5 Hz. Samples are each grown in 90 mTorr with temperature of 625, 635, 635, and 700 ºC for 0%, 10%, 30%, and 50% Sr respectively before being cooled in 200 T oxygen. X-ray diffraction [27] and electron microscopy [23] were used to confirm the single phase, cluster free nature of the films. After verification of sample quality, magnetometry measurements were performed using a Quantum Design MPMS3 magnetometer. All data are corrected for substrate background by subtracting the diamagnetic background signal. The contribution of the substrate to the magnetization is subtracted by measuring a SrTiO$_3$ substrate (from the same manufactured batch as those used to synthesize the samples) and directly subtracting the resulting signal scaled to the relative mass of the sample. In each case the substrate background was found exhibit only a diamagnetic signal, and therefore a linear subtraction is suitable. Subtraction of the vertical shift in the magnetization to determine the magnetiude of the pinned moment and exchange bias is done by taking the difference in the saturation magnetization of the two field biased states. An example of the subtraction is shown in Fig. 6 for the OOP FC measurements at 2.1 K.

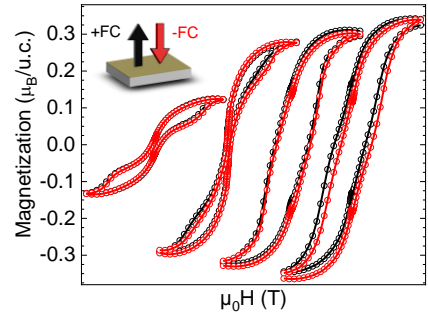

Fig. 6 Example showing the subtraction of the vertical shift for FC data at 2.1 K. This subtraction was used to determine the pinned moment and exchange bias reported in Fig. 3.

## Acknowledgements

All aspects of this work were supported by the US Department of Energy (DOE), Office of Basic Energy Sciences (BES), Materials Sciences and Engineering Division. Dynamic magnetometry measurements were performed as a user project at the Center for Nanophase Materials Sciences, which is sponsored at Oak Ridge National Laboratory (ORNL) by the Scientific User Facilities Division, BES, DOE. The work at Los Alamos National Laboratory was supported by the NNSA's Laboratory Directed Research and Development Program, and was performed, in part, at the CINT, an Office of Science User Facility operated for the U.S. Department of Energy

Office of Science through the Los Alamos National Laboratory. Los Alamos National Laboratory is operated by Triad National Security, LLC, for the National Nuclear Security Administration of U.S. Department of Energy (Contract No. 89233218CNA000001). Work at the National High Magnetic Field Laboratory is supported by NSF Cooperative Agreement No. DMR-1644779, the State of Florida and the US DOE. JS acknowledges support from the DOE Basic Energy Science Field Work Project Science in 100 T.